\documentclass[useAMS,usenatbib,onecolumn]{mn2e}

\usepackage{dcolumn}
\usepackage{graphicx}
\usepackage{url}
\usepackage{color}

\newcommand{\cm}{cm$^{-1}$}

\newcommand{\eqref}[1]{(\ref{#1})}

\title{ExoMol line lists III: An improved hot rotation-vibration line list for HCN and HNC}

\date{\today} 

\author[Barber et al]{R.J. Barber$^{1}$, J.K. Strange$^{1}$, C. Hill$^{1}$, O.L. Polyansky$^{1}$, G.Ch. Mellau$^{2}$,\newauthor  S.N. Yurchenko$^{1}$, Jonathan
Tennyson$^{1}$ \\
$^{1}$ Department of Physics and Astronomy, University College London, London WC1E 6BT, UK \\ 
$^{2}$ Physikalisch-Chemisches Institut, Justus-Liebig-Universit\"at Giessen, Heinrich-Buff-Ring 58, D-35392 Giessen, Germany}
\date{Accepted XXXX. Received XXXX; in original form XXXX}

\pagerange{\pageref{firstpage}--\pageref{lastpage}} \pubyear{2013}

\begin{document}

\maketitle

\begin{abstract}

  A revised rotation-vibration line list for the combined hydrogen
  cyanide (HCN) / hydrogen isocyanide (HNC) system is presented. The
  line list uses {\it ab initio} transition intensities calculated
  previously (Harris et al., ApJ, 2002, 578, 657) and extensive
  datasets of recently measured experimental energy levels (Mellau, J. Chem. Phys. and J. Mol. Spectrosc. 2010-2011). The
  resulting line list has significantly more accurate wavelengths than
  previous ones for these systems. An improved value for the
  separation between HCN and HNC is adopted leading to an
  approximately 25~\%\ lower predicted thermal population of HNC as a
  function of temperature in the key 2000 to 3000~K
  region. Temperature-dependent partition functions and equilibrium
  constants are presented. The line lists are validated by comparison
  with laboratory spectra and are presented in full as supplementary
  data to the article and at \url{www.exomol.com}.

\end{abstract}

\label{firstpage}

\section{Introduction}

Hydrogen cyanide (HCN) and its higher energy isomer hydrogen
isocyanide (HNC) are well known in a number of astrophysical
environments. For example, they are both prevalent in interstellar
clouds where HNC, despite lying at much higher energy, is often
observed to be of similar abundance as HCN \citep{98HiYaMi.HCN}.  HCN
and HNC are also observed well-outside thermodynamic equilibrium in
cometary atmospheres \citep{99HiYaKa.HCN} for which detailed
spectroscopic models are being developed \citep{13LiViDi.HCN}. 
Non-thermal HCN is also found on Titan where it is an important
coolant \citep{13ReKuFa.HCN}.
In cool
stars, where the two isomers exist in thermodynamic equilibrium, it
has been suggested their abundance ratio should act as a thermometer
\citep{jt321}.

In fact the opacity of HCN has a dramatic influence on the atmospheres
of cool carbon stars. \citet{84ErGuJo.HCN} showed that the inclusion
of an HCN line list \citep{85JaAlGu.HCN} resulted in models for the
atmospheres of cool carbon-rich stars in which the modelled atmosphere
expanded by a factor of five and the gas pressure of the surface
layers was lowered by one or two orders of magnitude. This clearly
illustrates the importance of accurate HCN opacities for such
models. Similarly, HCN spectra have been observed in T Tauri stars
\citep{11CaNa.HCN}, extra-galactic red giants \citep{06VaMaCo.HCN} and
have been used to constrain C and N abundances in AGB stars
\citep{mzv05}.

Neither HCN nor HNC has so far been detected in the atmospheres of
exoplanets. However HCN is predicted to be an important species and a
possible sighting has recently been claimed \citep{13OpBaBe.exo}. HCN
is likely to feature in carbon rich atmospheres, especially so when
disequilibrium effects are taken into account
\citep{12Madhux,12VeHeAg.exo,13MoKaVi}.  Indeed, surprise has been
expressed that HCN has yet to be firmly observed in exoplanets
\citep{13Moses}.  HCN is thus a key species for proposed exoplanet
characterisation studies \citep{jt523}. This work, along with the
other studies cited above, requires accurate spectroscopic data for
the HCN system.

\citet{84ErGuJo.HCN} computed a comprehensive vibration-rotation line
list using an {\it ab initio} quantum chemical procedure
\citep{85JaAlGu.HCN}. This line list was an important step forward but
is not particularly accurate and did not include any contribution from
HNC.  \citet{jt298} computed a new and even more extensive
vibration-rotation line list which covered both HCN and HNC. This work
was again based on {\it ab initio} quantum chemistry \citep{jt273}
which was shown to reproduce laboratory spectra with reasonable
accuracy \citep{jt283}. However, this line list, designated HPT below,
is far from spectroscopic accuracy. \citet{jt374} tried to improve the
accuracy of the line list by shifting the calculated energy levels to
ones measured experimentally, a procedure similar to that employed
below. However this effort was hampered by the lack of laboratory
data, particularly for levels important when the system is hot, so
that only 5200 energy levels were shifted. \citet{jt447} used a
similar technique to generalise the HPT line list to the H$^{13}$CN
system.

Since the work of \citet{jt374}, Mellau has performed systematic
experimental studies on the hot emission spectra of both HCN
\citep{08MeWiWi.HCN,11Mexxxx.HCN,11Mexxxb.HCN,11Mexxxc.HCN} and HNC
\citep{10Mexxxa.HNC,10Mexxxb.HNC,11Mexxxx.HNC}. This has led to the
experimental determination of over 40,000 rotation-vibration energy
levels of this system. Experimental energy level lists are provided in
the Supplementary Material of these papers
\citep{10Mexxxa.HNC,10Mexxxb.HNC,11Mexxxx.HCN,11Mexxxb.HCN,11Mexxxc.HCN,11Mexxxx.HNC}.
Up to 7000 cm$^{-1}$ above the HCN ground state and for 4000 cm$^{-1}$
above the HNC ground state Mellau's energy level list is complete and
high-resolution accurate. This complete list gives the first
practically error-free partition function at room temperature for a
polyatomic molecule. For the range of 7000 to 9000 cm$^{-1}$ above the
HCN ground state and 4000 to 6000 cm$^{-1}$ above the HNC ground state
a complete preliminary energy level list was computed by Mellau. The
energy level list in this region combines the experimental energy
levels with predicted energy levels. For the vibrational states not
covered by measurements, accurate global predictions of the
spectroscopic constants have been used to calculate the energy levels.
For each vibrational state the experimental energy levels are directly
related to measured transitions only for a given $J_{max}$. The energy
levels for each vibrational state have been extended up to $J=90$ for
HCN and up to $J=70$ for HNC with calculated energy levels based on
the spectroscopic constants determined using transitions up to
$J_{max}$.

In this work we merge Mellau's experimental and predicted energy level list with a modified HPT line list corrected for a more accurate position of the HNC minimum to provide a substantially improved single combined line list. Based on this data we give improved partition functions for high temperature applications.

The relative accuracies of energy level measurements are usually so
high that even if the relative error of the {\it ab initio} energy
levels are of the order of $10^{-3}$ to $10^{-4}$, they are many
orders of magnitude worse than experimental data. In the case of line
intensities, high-resolution accuracy means experimental line
intensities accurate to only to 1 to 10 \%. Due to this discrepancy
intensities derived from {\it ab initio} calculations are much more
comparable with measurements than is the case for the {\it ab initio}
frequencies. This work relies heavily on the assumption that {\it ab
  initio} transition intensities computed in the original HPT line
list are reliable. Mellau's group have performed extensive band
intensity analysis in the $\nu_1$ frequency region based on intensity
calibrated emission spectra. In this study they found an excellent
agreement between the measured and HPT line intensities up to a
relatively high bending excitation of $v_2=7$. This conclusion is
supported by comparisons with Mellau's laboratory emission spectra in
the $\nu_2$ bending region we give below, and by experience with
similar calculations on other molecules \citep{jt156,jt509}.

The current work forms part of the ExoMol project. This project aims
to provide line lists of spectroscopic transitions for key molecular
species which are likely to be important in the atmospheres of
extrasolar planets and cool stars; its aims, scope and methodology are
summarised in \citet{jt528}. This paper is third in series of
line list studies performed as part of this project. The previous
studies  concerned the diatomic species BeH, MgH and
CaH \citep{jt529}, and SiO \citep{jt563}.

\section{Method}

\subsection{Spectroscopic data}

The \citet{jt298} (HPT) line list consists of two files, one containing the energies of the eigenstates and their rotational and vibrational quantum numbers, where known (the states file), and the other a list of allowed transitions between these states (the transitions file). The list with the energies of the eigenstates contains 168,110 entries, of which 22,702 have ro-vibrational assignments. HPT considers rotationally excited states with $J$ up to 60.

Mellau's HCN experimental data set contains the eigenenergy for 26,344
eigenstates, of which 14,690 (55.8\%) are `$e$' parity and 11,654 are
`$f$' parity. These states correspond to 218 vibrational `$e$' or
`$f$' sublevels determined from the analysis of the emission spectra
and to 82 `$e$' or `$f$' sublevels calculated using predicted
spectroscopic constants. All sets have been extended using
spectroscopic predictions from the $J_{max}$ reported
\citep{08MeWiWi.HCN,11Mexxxx.HCN,11Mexxxb.HCN,11Mexxxc.HCN} up to
$J=90$. The HNC data set contains 14,284 HNC eigenstates, of which
8,136 (57.0\%) are `$e$' parity and 6,148 are `$f$' parity. The HNC
data set has 210 sets of vibrational `$e$' or `$f$' sublevels; each
set is extended from the reported $J_{max}$ value
\citep{10Mexxxa.HNC,10Mexxxb.HNC,11Mexxxx.HNC} up to $J=70$.  

As we place HCN and HNC levels on a single energy scale it is necessary
to align them both by allowing for the energy of moving from HCN to HNC
and by using a common cut-off for the levels of the two isomers. The
energy of the highest assigned HNC level in Mellau's experimental data
is at 16,000 \cm\ above the HCN ground state. As the highest assigned
HNC level in the HPT list is 10,000 \cm\ below this limit, only 
HNC levels in the experimental list with an energy less than 6,000
\cm\ above the HNC ground state were considered in this work.
This reduced the number of experimental HNC data to be examined to
2,001 `$e$' states and 1,471 `$f$' states. That is to say, of
approximately 30,000 experimental energy levels examined about 88\%\
were for HCN and 12\%\ for HNC. Moreover, relative to their respective
ground state energies, the HNC levels were at lower energies. It is
for this reason that in our reconstructed line list 15,201 HCN states
were substituted, but only 3,979 HNC states.

HPT took the isomerisation energy from HCN to HNC from
the  calculations of \citet{jt273} which places the HNC ground state 
5185.637 \cm\ above than of HCN.  At present there is only a highly
inaccurate experimental determination of the separation between the
HCN and HNC ground states \citep{82PaHe.HCN}, so Mellau's data gives
the energies of the HNC states relative to the HNC ground state.
Hence in order to make experimental HNC energies comparable with those
in the HPT list it is necessary to increase them by the energy
difference between the HNC and HCN ground states.  More recent
calculations suggest the value for this difference computed by
\citet{jt273} should be somewhat higher \citep{09DaWaTh.HCN}.  We
therefore undertook new calculations at the all-electron
multi-reference configuration interaction (MRCI) level of theory using
a large basis set (aug-cc-pCV6Z) and an extended active space for the
electron correlation; relativistic and adiabatic corrections were also
included. Full results of these calculations will be presented
elsewhere. These studies give an isomerisation energy of $5705 \pm 20$
\cm, including allowance for the zero point energy
corrections. 

The energies of the HNC levels in HPT were increased by 5705 --
5185.637 = 519.363 \cm\, and those from experimental HNC data by 5705 \cm. The
HNC ground state correction can be made only for the rotational
manifold of the first 40 HNC vibrational states labelled as HNC isomer
in the HPT list. The correction has consequences for the low and
medium temperature partition function of the system and the HCN/HNC
equilibrium value which are discussed below. In an overall HCN/HNC
opacity simulation using the corrected line list some minor artefacts are
expected for HNC transitions between corrected and uncorrected levels.

\subsection{The new energy file}

Due to the difference in the accuracy of the {\it ab initio} and measured eigenenergy data sets the bijective mapping of these data sets is straightforward only up to the fundamental vibrational excitation. Even at relatively low vibrational excitation it becomes increasingly complicated to map these data sets. As the density of states becomes an order of magnitude higher than the difference between measured and calculated eigenenergy values such mapping can be done only by labelling the {\it ab initio} eigenenergies with approximate vibrational quantum numbers and isomer labels. From the 168,110 eigenenergies of the HPT list 18,723 are identified as HCN and 3,979 as HNC. The remaining 145,408 states have rotational quantum numbers and parities, but the vibrational quantum numbers and the isomeric form are not known. 56.4\%\ of all the states identified are `$e$' parity and 43.6\%\ are `$f$' parity. The assignment was done before Mellau's list became available based on the sparse experimental data available at that time and on the rovibrational structure of the eigenenergies typical for a linear molecule. Mellau's measured and predicted list covers most of these assigned states. To check for possible errors in the original HPT assignment an experimental to {\it ab initio} mapping procedure was used in this work.   

The isomer code (0 for HCN, 1 for HNC), the parity (1 for `$e$' and 0
for `$f$' states) and the rotational and vibrational quantum numbers for
each entry in HPT states file were concatenated and tabulated against
the entry number in the file and the energy of the state.  Similarly
the experimental data which are organised by vibrational quantum
number were rearranged to match the HPT layout, and the isomer code,
parity, and rotational and vibrational quantum numbers were
concatenated and tabulated against energy. Matching entries were 
identified by using a simple look-up technique
that enabled the experimental energies for the corresponding states to
be matched against the HPT entries.  The data were then re-arranged
according to vibrational quantum numbers and in ascending $J$ values
within each vibrational state. Finally the difference between the HPT
energy and experimental energy was computed for each entry.

These differences found were systematic and made it easy to predict the energy
of the next highest $J$ with a high degree of accuracy (typically to
within 0.1 \cm). This enabled entries that had been wrongly assigned
in the HPT list to be readily identified.  Moreover, since the $J$
values are known to be correct and the energies of the state
corresponding to the wrongly assigned vibrational quantum numbers must
be in the same energy region, by sorting all the levels that had been
identified as mis-labelled by $J$ and then by energy, it was easy to
identify pairs of eigenstates whose quantum numbers had been wrongly
assigned.  Sometimes the confusion was identified as being between
three states with similar energies and on several occasions four
nearby states were found to all have quantum numbers belonging to
other members in the group. We found 302 HCN states that have
been wrongly assigned in HPT.  For values of $J<8$ no wrong
assignments were identified.  In the range $J=8-23$ and $J=50-60$ four
or fewer states were found to be wrongly assigned in the HPT list for
each value of $J$, whilst in the region $J=24-49$, with the exception
of $J=27$ (where there were only 4 wrong assignments), each value of
$J$ had between 6 and 13 states that were wrongly assigned. In the case of HNC, 
which as mentioned included far fewer states, no
meaningful analysis of the mis-assignments was possible, other than to
point out that all the wrongly labelled states were of `$f$' parity.

Next we substituted the experimental energies into the HPT list by
matching the quantum numbers.  In total, the energies of
15,201 HCN states were replaced with experimental energies, and of
these 8,544 were `$e$' parity and 6,657 `$f$' parity. The energies of
3,979 HNC states were replaced with experimental energies, of which
2325 were `$e$' states and 1,654 `f' states.

An examination of the HPT and experimental data revealed that the principal source of discrepancy between the two sets of data related to the vibrational band origins. Table~\ref{tab:obsdata} details 146 vibrational band origins for the HCN species, arranged by reducing value of the parameter $E_{\rm Expt}$ - $E_{\rm HPT}$. The difference in band origins has a strong vibrational angular momentum dependence as shown in Figure 2 of reference \citet{11Mexxxx.HCN}. The 21 bands showing the
greatest difference between the HPT and Mellau energies are highly excited bends with the vibrational angular momentum $l$ values very much lower than quanta of bending vibration, $v_2$.

Within each band there was also a
small systematic change (usually an increase) in the difference
between the two sets of energy with increasing values of $J$. However,
the differences due to small changes in $J$ were generally of second
order compared to the differences in band origins.

The HPT line list contains the rovibrational assignment of 200 vibrational states, this is 8 \% of the states below the first isomerisation state. Mellau assigned the rotational manifolds for more than 4000 vibrational states of the HPT line list based on spectroscopic models \citep{11Mexxxx.HCN}. Mellau's assignments agree for the first 71 vibrational states with the assignments reported in this work for all rotational states up to $J=60$. For each of these states the experimental eigenenergies have been reported up to $J=60$ \citep{11Mexxxx.HCN} and are included in the line list reported in this work. This means that using the line list reported in this work it is possible to make highly accurate simulations for \emph{any} band between these 71 vibrational states up to $J=60$. For the next 130 vibrational states there are differences between these two assignments mainly for highly excited rotational states with $J>35$. Altogether 5000 of the HPT states have different vibrational label in Mellau's assignment. The band centres of the two assignments match for all 265 different vibrational states including the further 65 highly excited vibrational states assigned only up to $J=5-15$ in the HPT line list.

{
\renewcommand{\thefootnote}{\alph{footnote}}
\renewcommand{\footnoterule}{\rule{0cm}{0cm}}

\begin{table}
\caption{Differences in the band origins between the experimental (Expt) values of Mellau and
HPT line list, given as observed minus calculated (o--c); all values in \cm.}
\label{tab:obsdata} \footnotesize
\begin{center}
\begin{tabular}{lrrrrrrlrrrrrrlrrrrrr}
\hline
Expt & $v_1$& $v_2$ &$\ell$& $v_3$&o--c&& 
Expt & $v_1$& $v_2$ &$\ell$& $v_3$&o--c&& 
Expt & $v_1$& $v_2$ &$\ell$& $v_3$&o--c\\
\hline
  6519.610  & 2& 0& 0& 0& +6.11&& 4994.340  & 0& 7& 5& 0& -3.43&& 6469.266  & 0&  9& 7& 0& -10.40\\
  9030.994  & 1& 8& 8& 0& +4.62&& 2893.589  & 0& 4& 4& 0& -3.46&& 8681.197  & 1&  8& 0& 0& -10.47\\
  8296.128  & 1& 7& 7& 0& +4.00&& 1411.413  & 0& 2& 0& 0& -3.50&& 3525.720  & 0&  2& 2& 1& -11.00\\
  3311.477  & 1& 0& 0& 0& +3.73&& 5580.396  & 0& 5& 1& 1& -3.54&& 4248.567  & 0&  3& 3& 1& -11.26\\
  6712.451  & 1& 5& 1& 0& +3.33&& 5414.442  & 1& 3& 3& 0& -3.63&& 8144.625  & 1&  1& 1& 2& -11.26\\
  6757.336  & 1& 5& 3& 0& +3.29&& 9258.983  & 2& 1& 1& 1& -3.64&& 5550.442  & 0&  8& 2& 0& -11.37\\
  6036.960  & 1& 4& 0& 0& +3.24&& 4888.039  & 0& 4& 0& 1& -3.72&& 7800.933$^1$& 0&  8& 6& 1& -11.91\\
  9190.518  & 2& 4& 2& 0& +3.14&& 2096.846  & 0& 0& 0& 1& -3.74&& 5525.813  & 0&  8& 0& 0& -11.95\\
  6060.819  & 1& 4& 2& 0& +3.00&&  714.936  & 0& 1& 1& 0& -3.86&& 6270.582  & 0&  3& 1& 2& -12.30\\
  9167.088  & 2& 4& 0& 0& +3.00&& 9459.164  & 1& 6& 2& 1& -4.01&& 9469.617  & 1&  9& 5& 0& -13.56\\
  5881.617  & 0& 8& 8& 0& +2.98&& 8563.197  & 2& 3& 3& 0& -4.07&& 4881.210  & 0&  1& 1& 2& -13.61\\
  5125.394  & 0& 7& 7& 0& +2.59&& 5625.238  & 0& 5& 3& 1& -4.16&& 6228.598  & 0&  0& 0& 3& -13.82\\
  7567.112  & 1& 6& 6& 0& +2.28&& 8034.720  & 1& 7& 1& 0& -4.30&& 7215.971$^1$& 0& 10& 8& 0& -14.03\\
  6644.127  & 0& 9& 9& 0& +2.18&& 4911.832  & 0& 4& 2& 1& -4.45&& 7770.008$^1$& 0&  5& 5& 2& -14.13\\
  7195.678  & 2& 1& 1& 0& +2.13&& 9435.375  & 1& 6& 0& 1& -4.74&& 6932.932  & 0&  1& 1& 3& -14.48\\
  7459.459  & 1& 6& 4& 0& +1.57&& 7443.402  & 1& 3& 1& 1& -5.10&& 7691.774$^1$& 0&  8& 4& 1& -14.62\\
  2802.959  & 0& 4& 0& 0& +1.50&& 4905.650  & 0& 7& 3& 0& -5.15&& 5571.734  & 0&  2& 0& 2& -14.77\\
  3498.086  & 0& 5& 1& 0& +1.27&& 6452.263  & 0& 6& 6& 1& -5.16&& 6336.631  & 0&  9& 5& 0& -14.96\\
  4375.376  & 0& 6& 6& 0& +1.15&& 6086.264  & 1& 1& 1& 1& -5.24&& 9381.428$^1$& 1&  9& 3& 0& -15.37\\
  7393.467  & 1& 6& 2& 0& +0.99&& 2161.615  & 0& 3& 3& 0& -5.29&& 6936.836  & 0&  7& 1& 1& -15.60\\
  8585.581  & 2& 0& 0& 1& +0.99&& 8880.764  & 1& 8& 6& 0& -5.31&& 7624.871$^1$& 0&  8& 2& 1& -16.07\\
  3543.625  & 0& 5& 3& 0& +0.93&& 6344.487  & 0& 6& 4& 1& -5.62&& 7039.129$^1$& 0&  4& 4& 2& -16.39\\
  2827.134  & 0& 4& 2& 0& +0.89&& 1435.440  & 0& 2& 2& 0& -5.68&& 9335.679$^1$& 1&  9& 1& 0& -16.56\\
  5369.820  & 1& 3& 1& 0& +0.75&& 6278.422  & 0& 6& 2& 1& -5.80&& 7600.536$^1$& 0&  8& 0& 1& -16.71\\
  8519.393  & 2& 3& 1& 0& +0.45&& 4859.669  & 0& 7& 1& 0& -6.08&& 5594.851  & 0&  2& 2& 2& -16.87\\
  7369.444  & 1& 6& 0& 0& +0.26&& 7455.423  & 1& 0& 0& 2& -6.15&& 6246.878  & 0&  9& 3& 0& -16.87\\
  7412.993  & 0&10&10& 0& +0.21&& 6254.406  & 0& 6& 0& 1& -6.18&& 6314.119  & 0&  3& 3& 2& -17.59\\
  4007.098  & 1& 1& 1& 0& -0.08&& 4204.148  & 0& 3& 1& 1& -6.37&& 7968.884$^1$& 0& 11& 9& 0& -17.61\\
  6843.879  & 1& 5& 5& 0& -0.08&& 8910.898  & 1& 5& 5& 1& -6.49&& 6200.333  & 0&  9& 1& 0& -17.84\\
  4265.998  & 0& 6& 4& 0& -0.68&& 8969.988$^1$& 0&12&12& 0& -6.71&& 7061.078$^1$& 0& 10& 6& 0& -19.96\\
  5393.698  & 1& 0& 0& 1& -0.73&& 5728.750  & 0& 8& 6& 0& -6.78&& 8728.056$^1$& 0& 12&10& 0& -21.13\\
  2116.414  & 0& 3& 1& 0& -0.85&& 5711.723  & 0& 5& 5& 1& -7.49&& 6949.074$^1$& 0& 10& 4& 0& -22.34\\
  3631.473  & 0& 5& 5& 0& -1.06&& 6760.705  & 1& 2& 0& 1& -7.80&& 6880.434$^1$& 0& 10& 2& 0& -23.62\\
  4198.983  & 0& 6& 2& 0& -1.23&& 8772.080  & 1& 8& 4& 0& -8.02&& 6855.443  & 0& 10& 0& 0& -24.16\\
  8167.247  & 1& 7& 5& 0& -1.53&& 7069.720  & 0& 7& 5& 1& -8.35&& 7791.506$^1$& 0& 11& 7& 0& -24.74\\
  4174.609  & 0& 6& 0& 0& -1.63&& 2808.518  & 0& 1& 1& 1& -8.37&& 7656.971$^1$& 0& 11& 5& 0& -27.68\\
  8780.819  & 1& 5& 1& 1& -1.86&& 4173.071  & 0& 0& 0& 2& -8.38&& 7565.908$^1$& 0& 11& 3& 0& -29.15\\
  4684.310  & 1& 2& 0& 0& -1.97&& 7641.128  & 0& 5& 1& 2& -8.69&& 8527.968$^1$& 0& 12& 8& 0& -29.17\\
  8825.250  & 1& 5& 3& 1& -2.09&& 6784.172  & 1& 2& 2& 1& -8.82&& 7518.738$^1$& 0& 11& 1& 0& -29.86\\
  7876.854  & 2& 2& 2& 0& -2.17&& 8196.387  & 1& 4& 4& 1& -8.83&& 8370.673$^1$& 0& 12& 6& 0& -32.72\\
  8107.969  & 1& 4& 0& 1& -2.29&& 3502.121  & 0& 2& 0& 1& -8.87&& 9270.461$^1$& 0& 13& 9& 0& -33.26\\
  7853.511  & 2& 2& 0& 0& -2.32&& 6951.683  & 0& 4& 0& 2& -9.30&& 8256.910$^1$& 0& 12& 4& 0& -34.34\\
  6126.351  & 1& 4& 4& 0& -2.36&& 7685.158  & 0& 5& 3& 2& -9.56&& 8187.207$^1$& 0& 12& 2& 0& -35.29\\
  8131.572  & 1& 4& 2& 1& -2.62&& 6982.219  & 0& 7& 3& 1& -9.66&& 8161.892$^1$& 0& 12& 0& 0& -35.66\\
  7951.843$^1$& 0& 8& 8& 1& -2.73&& 8705.445  & 1& 8& 2& 0& -9.67&& 9090.108$^1$& 0& 13& 7& 0& -37.46\\
  8188.274  & 0&11&11& 0& -2.80&& 4977.196  & 0& 4& 4& 1& -9.79&& 8953.319$^1$& 0& 13& 5& 0& -39.30\\
  4708.059  & 1& 2& 2& 0& -2.99&& 5618.181  & 0& 8& 4& 0& -9.90&& 8860.739$^1$& 0& 13& 3& 0& -40.30\\
  8079.988  & 1& 7& 3& 0& -3.01&& 7487.510  & 1& 3& 3& 1& -9.91&& 8812.732$^1$& 0& 13& 1& 0& -40.83\\
  7198.969  & 0& 7& 7& 1& -3.41&& 6975.047$^1$& 0& 4& 2& 2& -10.11&\\                             
\hline
\end{tabular}
\end{center}
 \noindent
  $^1$ Calculated using predicted spectroscopic constants. The estimated error of the predicted states  is less than 1 \cm. 
\end{table}

}

\subsection{Partition function}

We recomputed the temperature dependent partition function for the two
forms of the molecule: HCN and HNC, and produced eighth-order polynomial
fits to the data. The results for both HCN and HNC (relative to its own ground state) are
very close to those in \citet{jt304}, and hence are not reproduced here in tabular form.

For ease of use we fitted our partition function, $Q$, to a series expansion
of the form used by \citet{jt263}:
\begin{equation}
\log_{10} Q(T) = \sum_{n=0}^8 a_n \left[\log_{10} T\right]^n \label{eq:pffit}
\end{equation}
with the values given in Table~\ref{tab:pffit}.

\begin{table}
\caption{Fitting parameters used to fit the partition functions,
see eq.~\ref{eq:pffit}.Fits are valid for temperatures above 300 K}
\label{tab:pffit} \footnotesize
\begin{center}
\begin{tabular}{lccc}
\hline
      &    Total &     HCN &  HNC$^1$ \\  
\hline
$a_0$ &      -559.993791894 &       -200.584273948  & 2842.40726627    \\ 
$a_1$ &      1208.58614639  &       395.680813523   & -5546.4528922    \\
$a_2$ &      -1078.94207151 &       -286.330387905  &  4478.67281811   \\ 
$a_3$ &      511.99404245   &       71.9543795015   & -1913.15384532   \\ 
$a_4$ &      -136.18147179  &       19.3205191562   & 455.743140053    \\ 
$a_5$ &      19.2556107068  &       -17.8035517288  & -57.362270648    \\ 
$a_6$ &      -1.12987082345 &       4.90507954635   & 2.978749909      \\
$a_7$ &              0      &       -0.624341529866 &        0         \\
$a_8$ &              0      &      0.0308707949056  &        0         \\
\hline                                               
 \end{tabular}     

 \noindent
  $^1$ Partition function given relative to the HNC ground state.
\end{center}
\end{table}

An important parameter in thermal environments, such as the
atmospheres of cool stars, is the equilibrium constant, $K(T)$,
between HCN and HNC. \citet{jt304} estimated this using their partition
function. Given the new value of for the energy difference between HCN
and HNC used here, one would expect this value to shift. We have
therefore re-calculated $K$ as a function of temperature using the HCN
and HNC partition functions given in Table~\ref{tab:pffit} and an
energy separation of $5705$~\cm. The results are given in
Table~\ref{tab:equilib_K}. The increased energy separation results in
a reduction in the proportion of HNC by about 25~\%\ in the key
temperature range of 2000 to 3000~K compared to the earlier value of
\citet{jt304}.

\begin{table}
\caption{Equilibrium constant $K$, for HCN $\leftrightarrow$ HNC as a
  function of temperature. The percentage of HNC present in a thermalised
  sample is also given.}
\label{tab:equilib_K} 
\begin{center}
\begin{tabular}{rcr}
\hline
       $T$~(K)  &  $K(T)$ & \% HNC\\
\hline
800 &	0.00005	 &	0.005 \\
850 &	0.00010	 &	0.010 \\
900 &	0.00018  &      0.018 \\
950 &   0.00029	 &	0.029 \\
1000&	0.00046  &	0.046 \\
1100&   0.00097  &	0.097 \\
1200&	0.00188	 &      0.19  \\
1300&   0.00323	 &	0.32  \\
1400&   0.00523  &      0.52  \\
1500&   0.00789	 &      0.78  \\
1600&	0.01136  &	1.12  \\
1700&	0.01570	 &	1.55  \\
1800&	0.02089	 &	2.05  \\
1900&   0.02712	 &	2.64  \\
2000&   0.03426	 &	3.31  \\
2100&   0.04237	 &	4.06  \\
2200&	0.05131	 &      4.87  \\
2300&   0.06117  &	5.75  \\
2400&	0.07175	 &	6.68  \\
2500&	0.08304  &	7.64  \\
2600&   0.09489  &	8.63  \\
2700&	0.10732	 &	9.64  \\
2800&	0.12012	 &	10.66 \\
2900&	0.13328  &	11.67 \\
3000&	0.14667  &	12.68 \\
3100&   0.16024  &	13.68 \\
3200&   0.17391	 &      14.65 \\
3300&	0.18759  &	15.60 \\
3400&	0.20124	 &	16.53 \\
3500&	0.21482	 &	17.43 \\
3600&   0.22825	 &      18.29 \\
3700&	0.24152	 &	19.12 \\
3800&	0.25458  &	19.93 \\
3900&	0.26740	 &	20.69 \\
4000&	0.27998  &	21.43 \\
\hline
\end{tabular}
\end{center}
\end{table}

\subsection{Line list calculations}
\begin{figure}
\begin{center}
\includegraphics{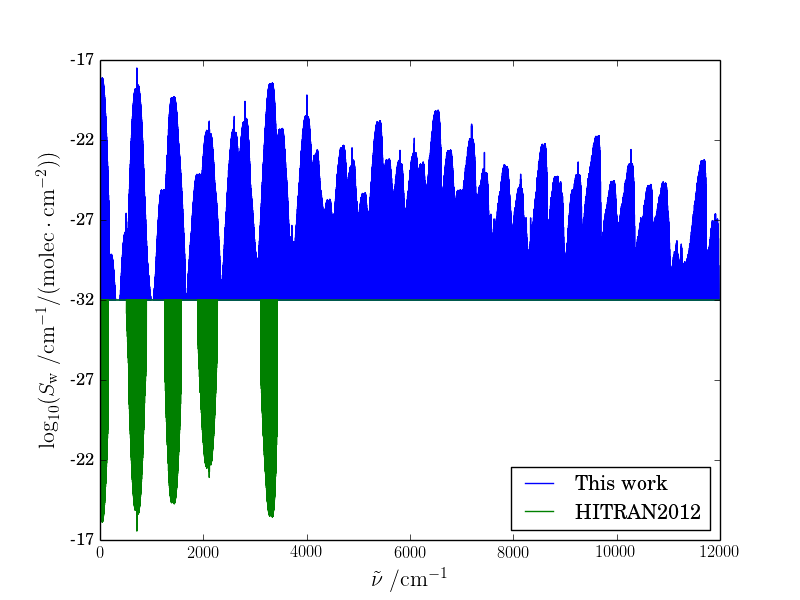}
\caption{Overview comparison of our line list with the data in HITRAN 2012 \citep{jt557}. }
\label{fig:HITRAN}
\end{center}
\end{figure}

In principle, having substituted the energy levels, the line list is
automatically generated using the transitions file. In practice one
more refinement needed to be made. The re-ordering of the energy
levels and the shift in the HNC ground state energy led a small
minority of transitions whose upper and lower state energies are now
inverted giving rise to unphysical transitions with negative
frequencies. Since all of these transitions are poorly characterised
ones between high energy states, they were simply removed from the
transitions file.

 %\subsection{Intensities}

\section{Results}
\begin{table}
\caption{Extract from the states file.}
\label{tab:levels} 
\begin{center}
\begin{tabular}{lccccccccc}
\hline
        $I$  &  $\tilde{E}$      &  $g$  &  $J$ & parity & iso & $v_1$ & $v_2$ & $l_2$ & $v_3$\\
\hline
            1 &    0.000000 &   6   &    0 & + & 0 &   0 & 0 & 0 & 0\\
            2 & 1411.413450 &   6   &    0 & + & 0 &   0 & 2 & 0 & 0\\
            3 & 2096.845540 &   6   &    0 & + & 0 &   0 & 0 & 0 & 1\\
            4 & 2802.958740 &   6   &    0 & + & 0 &   0 & 4 & 0 & 0\\
            5 & 3311.477080 &   6   &    0 & + & 0 &   0 & 1 & 0 & 0\\
            6 & 3502.121100 &   6   &    0 & + & 0 &   0 & 2 & 0 & 1\\

\hline
\end{tabular}

\noindent
  $I$:   State counting number;
 $\tilde{E}$: State energy in \cm;
  $g$: State degeneracy;
$J$:   State rotational quantum number;
parity: Total parity of the state: + or $-$;
iso: isomer label, 0 is HCN, 1 is HNC, 2 is unspecified;
 $v_1$:   $\nu_1$ state vibrational quantum number;
 $v_2$:   $\nu_2$ state vibrational quantum number;
 $l_2$:   vibrational angular momentum associated with $\nu_2$;
 $v_3$:   $\nu_3$ state vibrational quantum number.
\end{center}
\end{table}

\begin{table}
\caption{Extracts from the transitions file.}
\label{tab:trans} 
\begin{center}
\begin{tabular}{rrr}
\hline
       $I$  &  $F$ & $A_{IF}$\\ 
       31545   &    29398 & 7.7150e-10\\
      139218   &   136884 & 6.1040e-06\\
      112117   &   110541 & 1.8330e-04\\
      161475   &   160713 & 1.1600e-04\\
       76389   &    74341 & 1.0340e+01\\
       70455   &    74326 & 6.8550e-04\\
\hline
\end{tabular}

\noindent
 $I$: Upper state counting number;
$F$:      Lower state counting number;
$A_{IF}$:  Einstein A coefficient in s$^{-1}$.

\end{center}
\end{table}

Samples of the revised states and transitions files are given in
tables~\ref{tab:levels} and \ref{tab:trans} respectively. Full
versions of these files can be downloaded from the Strasbourg data
centre, CDS, via ftp://cdsarc.u-strasbg.fr/pub/cats/J/MNRAS/ or from
the ExoMol website, www.exomol.com.

Figure~\ref{fig:HITRAN} gives an overview of the line list at room
temperature, 296~K, in comparison with the, largely experimental, data
in HITRAN database \citep{jt557}.  HITRAN only contains HCN, which
itself is not an important atmospheric species. It is clear that
HITRAN does not give a complete representation of HCN spectral data
and, in particular, is missing bands in the region of 4~$\mu m$,
associated with the $\nu_1 - \nu_2$ difference band, at about 2600
\cm, and the $\nu_2 + \nu_3$ combination band, near 2800 \cm.

\begin{figure}
\begin{center}
\includegraphics[scale=0.75]{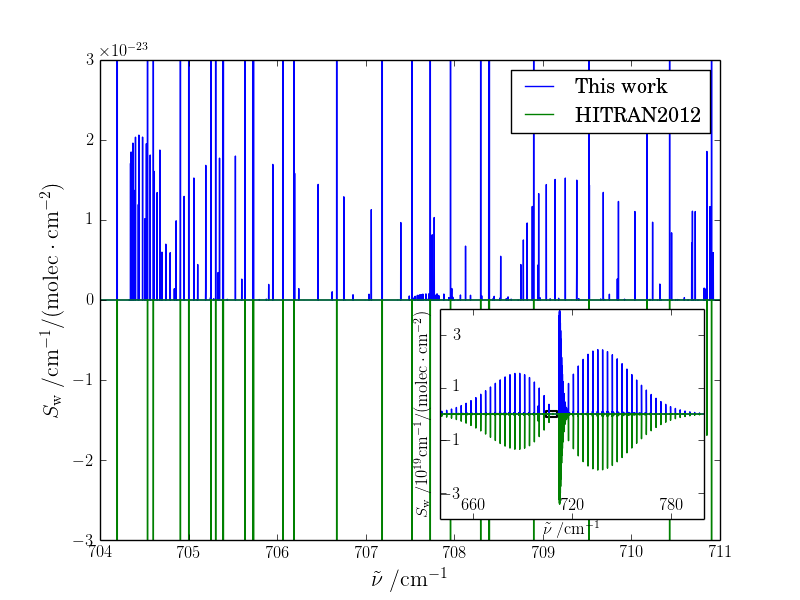}
\includegraphics[scale=0.75]{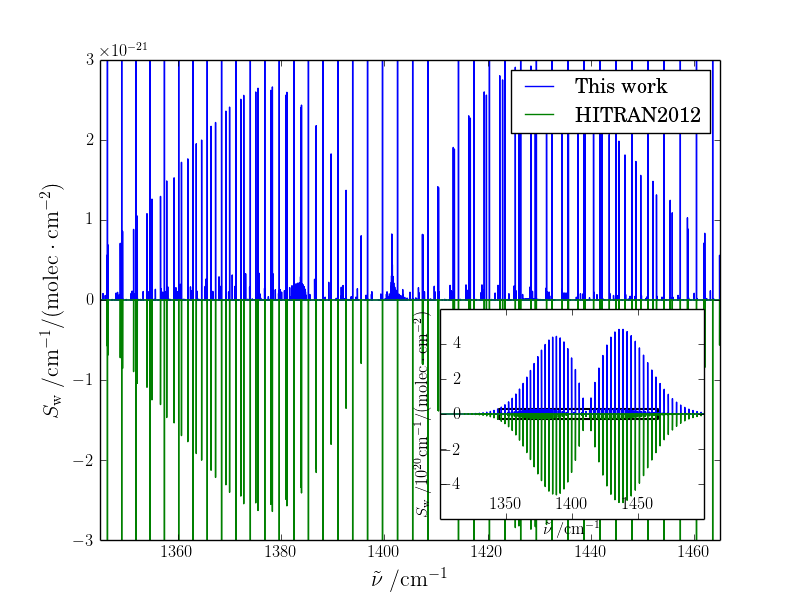}
\caption{Comparison of our line list with the data in HITRAN 2012 \citep{jt557}
for the region of the $\nu_2$ band (upper) and $2\nu_2$ band (lower). }
\label{fig:twov2}
\end{center}
\end{figure}

Figure~\ref{fig:twov2} gives an
expanded view of the regions covering the HCN bending fundamental, $\nu_2$, and overtone,
$2\nu_2$.  Unsurprisingly there is excellent agreement between the
frequencies in our new line list and those in HITRAN as they are based fully on Mellau's data. The agreement
between the line intensities is also excellent, proving again our conclusion regarding the accuracy of the {\it ab initio} intensities. Our line list
indicates the presence of many weak lines, which appear as ``grass'' in
the expanded scale spectrum in Figure~\ref{fig:twov2}, which are
absent from HITRAN. These lines can be associated with hot bands and
are significantly more intense at higher temperatures.

\begin{figure}
\begin{center}
\includegraphics{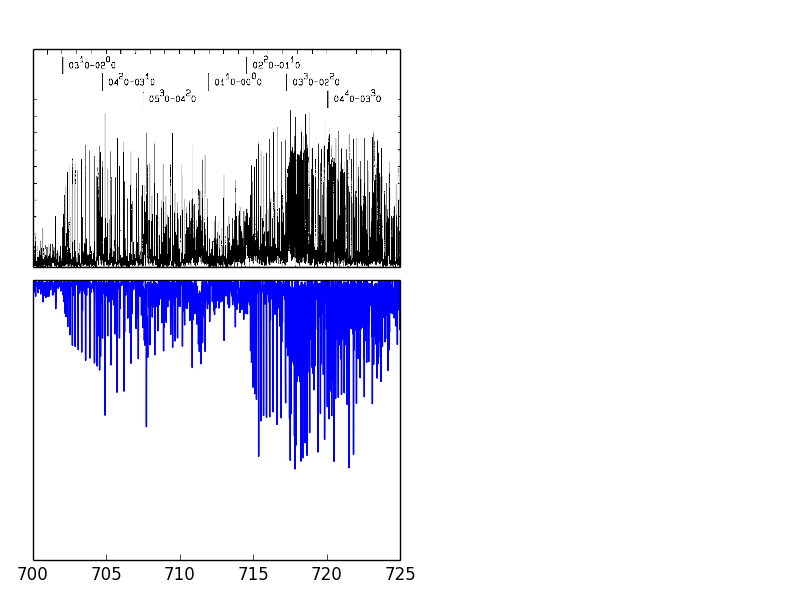}
\caption{Comparison of our line list with the emission spectra of \citet{00MaMeKl.HCN} assuming
a temperature of 1370~K. The first lines of several Q-branches are indicated at the top of 
the figure.}
\label{fig:Maki}
\end{center}
\end{figure}

While there are many laboratory studies of the spectra of room
temperature HCN and HNC, studies of hot spectra are much
rarer. Mellau's emission spectra do not contain absolute intensity
information and the line frequencies in our line list agree with these
experimental measurements by construction, so a comparison is not
informative. \citet{00MaMeKl.HCN} recorded an emission spectrum of HCN
at about 1370~K. The intensities in this spectrum are subject to
self-absorption by cooler HCN but a comparison, as given in
figure~\ref{fig:Maki} is still useful. Again, although intensities
disagree in details for the reasons given above, the overall structure
of the two spectra are very similar.

\section{Conclusions}

We present a new line list which should produce hot HCN and HNC
spectra with significantly improved wavelengths. In particular, the
Q-branches, which give sharp spikes in the spectrum, should now be
correctly located. As these features are likely to provide the key to
any detection of HCN in extrasolar planets, something that is very
much expected \citep{13Moses}, we believe the improved line list
should provide the necessary spectroscopic data.  The new line list
may be accessed via www.exomol.com or
http://cdsarc.u-strasbg.fr/viz-bin/qcat?J/MNRAS/.  The states and
transitions files may be used to compute temperature-dependent
spectra.  Alternatively, the ExoMol website provides a facility for
providing these data as temperature-dependent cross sections
\citep{jt542}.

Finally we note that the procedure used here can be employed to
improve the wavelengths in any of the line lists generated as part of
the ExoMol project.  The use of the MARVEL (Measured Active
Rotational-Vibrational Energy Levels) \citep{jt412} provides the means of
inverting laboratory measurements to give reliable energy levels which
can be actively improved as new studies become available. The results
of such MARVEL studies can then be used to update the relevant levels
file.  This procedure was used, at least in part, for water in the
latest release of HITEMP \citep{jt480}; and is likely to be used for
other line lists in the future.

\section*{Acknowledgements}

This work is supported by ERC Advanced Investigator Project 267219.
JS thanks the Nuffield Foundation for funding under the Nuffield
Research Placement Scheme.

\bibliographystyle{mn2e}
%\bibliography{journals_astro,exogen,jtj,hcn,methods,additional,linelists}

\label{lastpage}

\end{document}